\documentclass[fleqn,usenatbib]{mnras}

\usepackage{newtxtext,newtxmath}

\usepackage[T1]{fontenc}

\DeclareRobustCommand{\VAN}[3]{#2}
\let\VANthebibliography\thebibliography
\def\thebibliography{\DeclareRobustCommand{\VAN}[3]{##3}\VANthebibliography}

\usepackage{graphicx}	
\usepackage{amsmath}	

\usepackage{graphicx}
\usepackage{float}
\usepackage{subfigure}

\title[A multimodal machine learning for Photo-z]{PhotoRedshift-MML: a multimodal machine learning method for estimating photometric redshifts of quasars}

\author[Shuxin Hong et al.]{
Shuxin Hong,$^{1,2,3,4}$
Zhiqiang Zou,$^{1,2}$\thanks{E-mail: zouzq@njupt.edu.cn}
A-Li Luo,$^{3,4}$\thanks{E-mail: lal@nao.cas.cn}
Xiao Kong,$^{3}$
Wenyu Yang$^{1,2}$
and Yanli Chen$^{1,2}$
\\
$^{1}$College of Computer, Nanjing University of Posts and Telecommunications, Nanjing 210023, China\\
$^{2}$Jiangsu Key Laboratory of Big Data Security and Intelligent Processing, Nanjing 210023, China\\
$^{3}$CAS Key Laboratory of Optical Astronomy, National Astronomical Observatories, Beijing 100101, China\\
$^{4}$School of Astronomy and Space Science, University of Chinese Academy of Sciences, Beijing 100049, China
}

\date{Accepted XXX. Received YYY; in original form ZZZ}

\pubyear{2022}

\begin{document}
\label{firstpage}
\pagerange{\pageref{firstpage}--\pageref{lastpage}}
\maketitle

\begin{abstract}
We propose a Multimodal Machine Learning method for estimating the Photometric Redshifts of quasars (PhotoRedshift-MML for short), which has long been the subject of many investigations. Our method includes two main models, i.e. the feature transformation model by multimodal representation learning, and the photometric redshift estimation model by multimodal transfer learning. The prediction accuracy of the photometric redshift was significantly improved owing to the large amount of information offered by the generated spectral features learned from photometric data via the MML. A total of 415,930 quasars from Sloan Digital Sky Survey (SDSS) Data Release 17, with redshifts between 1 and 5, were screened for our experiments. We used |$\Delta$z| = $|(z_{phot}-z_{spec})/(1+z_{spec})|$ to evaluate the redshift prediction and demonstrated a $4.04\%$ increase in accuracy. With the help of the generated spectral features, the proportion of data with |$\Delta$z| < 0.1 can reach  $84.45\%$ of the total test samples, whereas it reaches $80.41\%$ for single-modal photometric data. Moreover, the Root Mean Square (RMS) of |$\Delta$z| is shown to decreases from 0.1332 to 0.1235. Our method has the potential to be generalized to other astronomical data analyses such as galaxy classification and redshift prediction.

\end{abstract}




\section{Introduction}



Quasars are extremely bright active galactic nuclei (AGN), one of the brightest, most powerful and most energetic objects known in the universe \citep[][]{peterson1997introduction}, and a critical class of samples for studying cosmology and galaxy physics. With the Wide Field Survey, more than 750,000 quasars have been discovered \citep[][]{lyke2020sloan}, most of them from the Sloan Digital Sky Survey \citep[SDSS,][]{york2000sloan}. According to Hubble's law, quasars that are generally farther away tend to have larger redshifts. To determine the redshift accurately, a spectroscopic observation is required. However, the magnitude limit in the telescope's observation spectrum is often lower than that of photometric observation, which necessitates the selection of a target. On the other hand, despite the existence of spectroscopic observations, not enough features are available in some particular wavelength's band to enable an accurate redshift determination. Therefore, obtaining a complete spectroscopic redshift of quasars in a short time remains challenging in this case. In contrast, the photometric redshift ($z_{phot}$) is inferred from the multi-band photometric data set, which greatly reduces the resources budget and can provide a relatively complete data sample for the study of galaxy formation, evolution and cosmology. However, increasing the precision of the photometric redshift measurement constitutes an essential requirement for this sort of investigations.

Various techniques are available to obtain photometric redshifts, ranging from template-fitting to machine learning (hereafter ML) and hybrid systems \citep[][]{salvato2019many}. Here, we focus on the latter two types of methods. At present, data mining and ML technologies are being widely used for dealing with massive astronomical spectroscopic and photometric data, owing to their high efficiency and accuracy \citep[][]{brescia2021photometric, fluke2020surveying, baron2019machine}. They can be divided into two categories: supervised and unsupervised learning. Supervised ML methods require training data with both photometric data as inputs and spectroscopic redshifts as labels. Gaussian Processes (GP) can be used to produce photo-z estimations \citep[][]{duncan2018photometric, ansari2021mixture}. Tree-based algorithms, such as decision trees and random forests, are often utilized to provide an estimation of photometric redshifts \citep[][]{carliles2010random, hoyle2015data, mountrichas2017estimating}. Support vector machine (SVM) \citep[][]{peng2010support} and K-Nearest Neighbour (KNN) \citep[][]{ball2008robust, curran2020qso} are also widely used for their simplicity. Neural Networks are popular supervised ML algorithms \citep[][]{bonnett2015using} including methods such as artificial neural networks (ANN) \citep[][]{yeche2010artificial, sadeh2016annz2}, deep neural networks (DNN) \citep[][]{hoyle2016measuring}, convolutional neural networks (CNN) \citep[][]{d2018photometric, pasquet2019photometric, mu2020photometric}, and many others. The Multi-Layer Perceptron with Quasi-Newton Algorithm (MLPQNA) is another very popular deep learning method for estimating photo-z \citep[][]{cavuoti2012photometric, cavuoti2017metaphor, brescia2019photometric, razim2021improving}, among which \citet{brescia2019photometric} computed photometric redshift for AGN and compared it to spectral energy distribution (hereafter SED) fitting technique using LEPHARE \citep[][]{arnouts1999measuring, ilbert2006accurate}. Moreover, the combination of various data mining technologies can constitute an interesting approach to obtain the photometric redshift of galaxies and quasars \citep[][]{laurino2011astroinformatics}. The aforementioned supervised ML methods have certain advantages, but they all depend on label data, in absence of which, unsupervised ML becomes a more interesting option. Unsupervised ML often uses clustering-based methods \citep[][]{rahman2015clustering, scottez2016clustering}, a most popular example is the Self Organising Map (SOM) \citep[][]{speagle2017deriving}.

Hybrid systems often combine template-fitting and machine learning methods \citep[][]{salvato2019many}. For example, \citet{carrasco2014exhausting} presented a novel and efficient Bayesian framework, which combines the results from different photo-z techniques, such as random forest, SOM, and a standard template-fitting method. \citet{beck2016photometric} used local linear regression method for the redshift and redshift error estimation, followed by a template-fitting step. \citet{leistedt2017data} combined the best of the ML and SED fitting techniques. Moreover, some researchers have reported comparative approaches between the aforementioned methods. \citet{desprez2020euclid} compared thirteen different photo-z methods, either template-fitting based or machine-learning based. \citet{schmidt2020evaluation} compared twelve template based and machine-learning based photo-z PDF codes and observed that no one code dominates in all metrics. For galaxy photometric redshift, the prediction results of the above methods are similar, and the error of their prediction results is relatively small.

Compared to galaxies, which are extended sources, quasars are points, inferring much lower amount of pixels captured by the CCD, which limits the precision of quasars photometry. On the other hand, the broad emission line of quasars introduces redshift uncertainty using photometric data. To improve the prediction accuracy of the quasars' photometric redshift, \citet{wu2010quasar} combined the optical data of SDSS and the near-infrared data of UKIDSS, which significantly improved the selection efficiency and photometric redshift accuracy of quasars. \citet{zhang2013estimating} combined SDSS, UKIDSS and WISE multi-band data, and used KNN to improve the prediction accuracy of quasars' photometric redshift. \citet{brescia2013photometric} combined data from four sky surveys (SDSS, GALEX, UKIDSS and WISE), covering a wide range of wavelengths from UV to mid-infrared, and used MLPQNA to obtain high photometric redshift prediction accuracy. The above research simply joins the features from different telescopes without mining the natural relationship between them. They depend on the high-quality data of multiple telescopes, which has the problem of data dependence. However, when there is only one kind of data from SDSS, the above methods will be invalid as they do not apply multi-modal deep mining to these data.

Multimodal machine learning (hereafter MML) \citep[][]{baltruvsaitis2018multimodal} has become a research hotspot in recent years for it is closer to the human learning style. \citet{wu2014exploring} fused image and audio data through multimodal representation learning and achieved a better classification effect compared to the case of single-modal data. \citet{mroueh2015deep} fused the mouth image data and sound data of human speech based on the multimodal deep learning model, which improved the recognition effect of mouth images. \citet{owens2016visually} learned the features of two modal data (video and audio) through CNN and Long Short Term Memory (LSTM) and successfully simulated the audio data from the video data. Transfer learning in multimodal machine learning is to use information from one resource-rich modality to assist another relatively resource-poor modality to learn, which can solve the problems of data dependence and so on. \citet{fu2021finding} considered the inconsistent data distribution of quasars in low silver and high Galactic latitudes and successfully constructed the missing quasar optical data in low silver latitude using the existing quasar optical data in high silver latitude based on the transfer learning.

Inspired by the idea of supervised MML, we introduce MML to represent the photometric and spectroscopic data of SDSS formally through transferring photometric data to the feature space of spectra and using the converted features to assist the prediction of quasars' photometric redshift. Specifically, the high-dimensional spectroscopic data are represented, and then the photometric magnitude is transferred to the spectral feature space. There are two kinds of data consisting of SDSS spectrum and photometry in the training samples, but only SDSS photometry data in the testing samples. In the pre-training stage, CNN with an attention mechanism is used to learn feature knowledge from high-dimensional spectroscopic data. Then, we build a generative model to train photometric data to generate simulated spectral data, which are very close to the true spectra, so as to assist the photometric data in the downstream task to complete the redshift prediction and improve its accuracy. To the best of our knowledge, few people have used the multimodal transfer learning method to predict the photometric redshift of quasars. The experimental results show that our method significantly improves the accuracy of photometric redshift estimation of quasars.

The main contributions of this study are as follows: (1) A photometric to spectral feature transformation model (i.e., PhotoSpecTransformer), based on multimodal representation learning, is constructed to represent data pairs (photometric data features, spectroscopic data features) in a unified feature space. The goal of the PhotoSpecTransformer model is to generate simulated spectral data from photometric data; (2) A photometric redshift estimation model based on transfer learning (i.e., PhotoReshift-TL) is constructed to assist the photometric data with the generated simulated spectral data, so as to improve the accuracy of redshift prediction; (3) Our method has generality to some extent. If the even pair data of quasars are replaced by that of galaxies, the redshift of galaxies can also be predicted with higher accuracy; (4)We share the codes and experimental data, published in GitHub (https://github.com/HongShuxin/PhotoRedshift-MML), to facilitate the access to their usage by other researchers to reproduce the research and further expand.

The remaining of this paper is organised as follows. In Section 2, we introduce the acquisition, composition, and pre-processing of data. In Section 3, we present the method used in this paper, its implementation process, and the structure of the models in detail. Section 4 gives the evaluation metrics, comparisons, and analyses of experimental results. Finally, Section 5 summarizes and discusses the key findings of our work.

\section{Data}
This section mainly introduces the relevant information of the data set we use in the present study. We first present the source and composition of the data set. Then, we describe a series of data pre-processing work, aiming at providing appropriate input data for subsequent experimental models.

\subsection{Data acquisition}
The photometric and spectroscopic data used in this study originate from the SDSS Data Release 17\footnote{https://www.sdss.org/dr17/}. Since multimodal learning requires paired photometric and spectroscopic data, it is necessary to select quasar photometric data with the corresponding observation spectra. We combined specObjAll and photoObjAll catalogues and used SQL statements to query the data on SDSS Casjob\footnote{http://skyserver.sdss.org/CasJobs/}. A total of 520,335 qualified quasar photometric data were retrieved. The SQL query used is listed in Appendix A.

\begin{figure}
\centering
\includegraphics[width=\hsize]{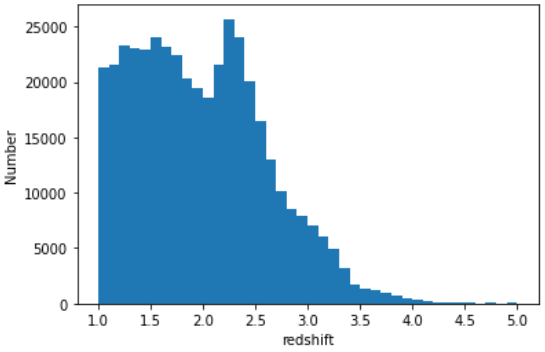}
\caption{Redshift distribution histogram.}
\label{f1}
\end{figure}

We selected the PSF (point spread function) magnitude (Fukugita et al., 1996)  of five bands: $u-$, $g-$, $r-$, $i-$, and $z-$band (denoted as $psfMag\_u$, $psfMag\_g$, $psfMag\_r$, $psfMag\_i$, $psfMag\_z$, respectively), the extinction of each band (denoted as $extinction\_u$, $extinction\_g$, $extinction\_r$, $extinction\_i$, $extinction\_z$, respectively), and errors of five bands (denoted as $psfMagErr\_u$, $psfMagErr\_g$, $psfMagErr\_r$, $psfMagErr\_i$, $psfMagErr\_z$). The redshift range is limited to $1\leq z\leq 5$, and the $g$ magnitude range is limited to $18\leq psfMag\_g\leq 22$, both according to \citet{yeche2010artificial}. The $snMedian$ represents the signal-to-noise ratio (SNR for short), we chose data with $snMedian$ < 10 in all bands, for the data with lower SNR needs to improve the prediction accuracy through multimodal methods. The $petroRad\_r$ should be less than 5, so that some of AGN can be eliminated from our quasar data. There are 520,335 photometric data meeting the SQL search conditions. In which, there will be duplicate sources, that is, data with the same objID, which were eliminated. Only the data with the smallest zErr is retained for each source, and finally 416,296 targets are obtained. The data volume distribution under different redshifts is shown in Figure~\ref{f1}.

Furthermore, the corresponding spectra of the 416,296 quasars are downloaded on the DR17 science archive server (SAS)\footnote{https://dr17.sdss.org/optical/spectrum/search} according to the identification of (plate, mjd, fiberId) triplet.

\subsection{Data pre-processing}
Data pre-processing aims at providing the model with reliable and high-quality data by eliminating the effect of skylight and dimension. In the following, we introduce the processing methods of photometric and spectroscopic data, respectively.

\subsubsection{Pre-processing of photometric data}
For photometric data, extinction correction should first be carried out in advance. The magnitude of each band should be subtracted from the extinction of the corresponding band, given by equation~(\ref{m1}) as:
\begin{equation}
    mag\_k = psfMag\_k - extinction\_k,
    \label{m1}
\end{equation}
where $k$ represents one of the five bands, $psfMag\_k$, $extinction\_k$, and $mag\_k$ represent the original magnitude, the extinction, and the magnitude after extinction correction of $k-$band, respectively.

Secondly, to eliminate the influence of distance, and  increase the number of features, we calculate the values of five colours, which are $mag\_u$-$mag\_g$, $mag\_g$-$mag\_r$, $mag\_r$-$mag\_i$, $mag\_i$-$mag\_z$, $mag\_u$-$mag\_z$, respectively. So far, we have got 15 dimensional photometric data, i.e., $mag\_u$, $mag\_g$, $mag\_r$, $mag\_i$, $mag\_z$, $mag\_u$-$mag\_g$, $mag\_g$-$mag\_r$, $mag\_r$-$mag\_i$, $mag\_i$-$mag\_z$, $mag\_u$-$mag\_z$, $psfMagErr\_u$, $psfMagErr\_g$, $psfMagErr\_r$, $psfMagErr\_i$, $psfMagErr\_z$, as the input features of the quasar redshift prediction model. 

Thirdly, before inputting into the model, the 15 features should be normalized for the sake of measurement's consistency. Considering that their numerical range is relatively concentrated, linear normalization is adopted. The calculation method is shown in equation~(\ref{m2}). The calculation results would build a 15-dimensional vector, denoted as $\chi_{phot}$.
\begin{equation}
    X_{j}\_norm=\frac{X_{j}-X_{j}\_min}{X_{j}\_max-X_{j}\_min},
    \label{m2}
\end{equation}
where $X_{j}$ represents the original value of the $j$-th feature ($1\leq j\leq 15$), $X_{j}\_min$ and $X_{j}\_max$ represent the minimum and maximum values of the $j$-th feature, respectively, and $X_{j}\_norm$ represents the $j$-th eigenvalue after normalization.

\subsubsection{Pre-processing of spectroscopic data}
For spectroscopic data, we should first ensure the consistency of input dimensions. Considering the inconsistency of skylight residues at the red end of spectroscopic data \citep[][]{r1, zou2020celestial}, the first 3600 dimensional features of spectroscopic data are uniformly retained. In this process, it is found that the dimension of a small number of spectra is less than 3600, so this part of spectra and their corresponding photometric data are removed from the data set. Further, there are 415,930 quasar data left from the original 416,296 quasars.

Secondly, spectroscopic data also need to be normalized. The flux at different wavelengths of the spectrum varies greatly. If not treated, the features of high intensity will dominate, whereas the low intensity features will be ignored by the model. To address this issue, we use the method called flux standardization \citep[][]{r2} to normalize the spectroscopic data, which could eliminate the influence of dimension and accelerate the convergence speed of the model. The specific calculation method can be seen in equation~(\ref{m3}). The calculation results can form a 3600-dimensional vector, denoted as $\chi_{spec}$.
\begin{equation}
    y_{m} = \frac{y_{m}}{\left \| y_{m} \right \|_{2}},
    \label{m3}
\end{equation}
where $y_{m}$ represents the $m$-th spectrum, $\left \| y_{m} \right \|_{2}$ represents the 2-norm of the $m$-th data obtained by computing the square root of the sum of squares of each element.

\section{Methods}
\begin{figure}
\centering
\includegraphics[width=\hsize]{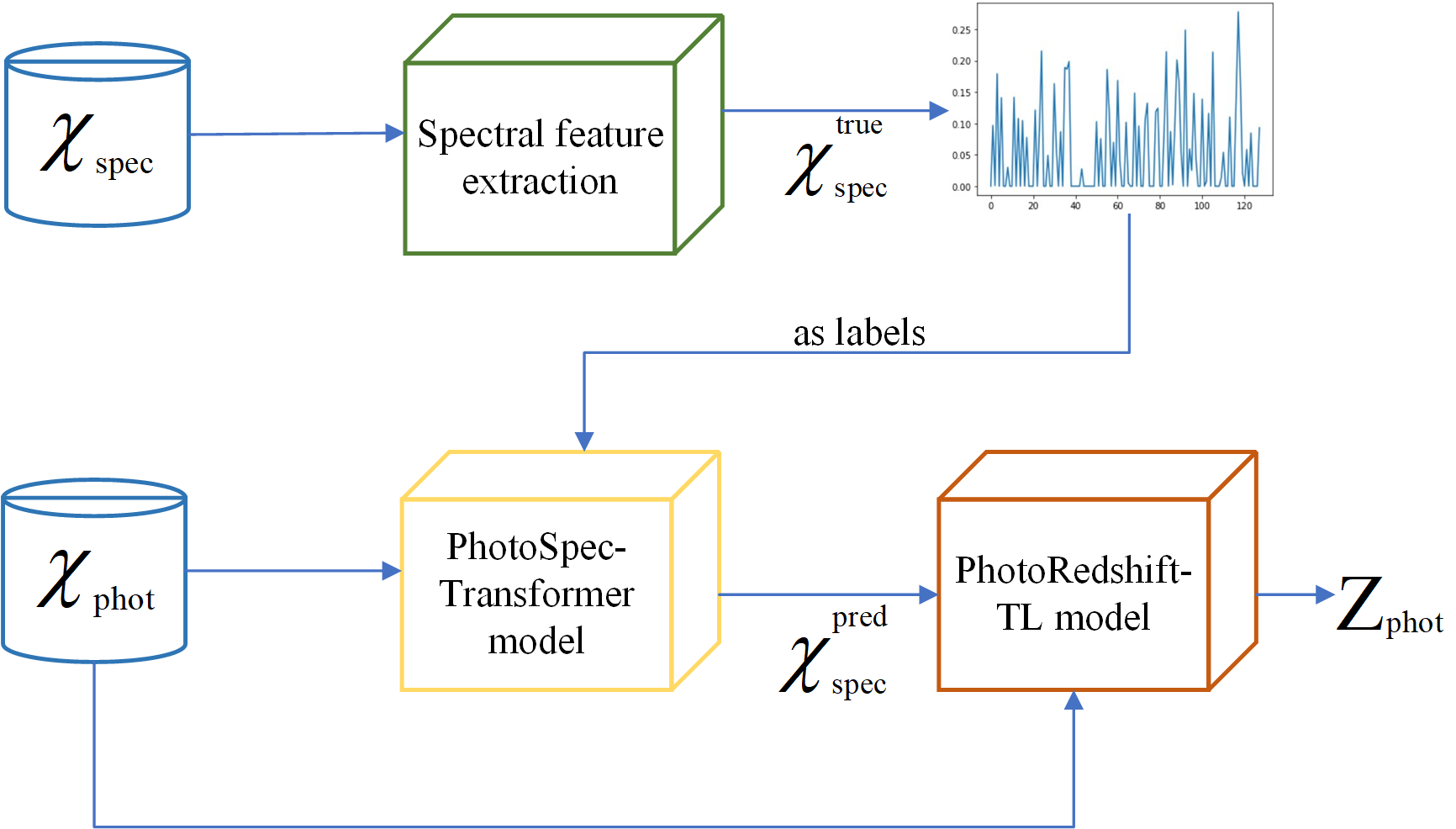}
\caption{Framework of PhotoRedshift-MML.}
\label{f2}
\end{figure}

Considering that the quasar photometric data contains too few features, which leads to low photometric redshift prediction accuracy, a Photometric Redshift estimation based on the Multimodal Machine Learning (PhotoRedshift-MML) method is proposed to improve the accuracy of photometric redshift prediction, as shown in Figure~\ref{f2}. PhotoRedshift-MML consists of two parts, one is a photometric to spectral feature transformation model (PhotoSpecTransformer), whose specific implementation will be introduced in Section 3.2.1. The other is a photometric redshift estimation model (PhotoRedshift-TL), whose specific implementation will be introduced in Section 3.2.2. In order to verify the advantage of the multimodal method over the single-modal method, we first constructed a photometric redshift prediction model based on single-modal machine learning in Section 3.1, which is used as one of the baselines of our work.

\subsection{Prediction of quasars' photometric redshifts based on single-modal machine learning}
\begin{figure}
\centering
\includegraphics[width=\hsize]{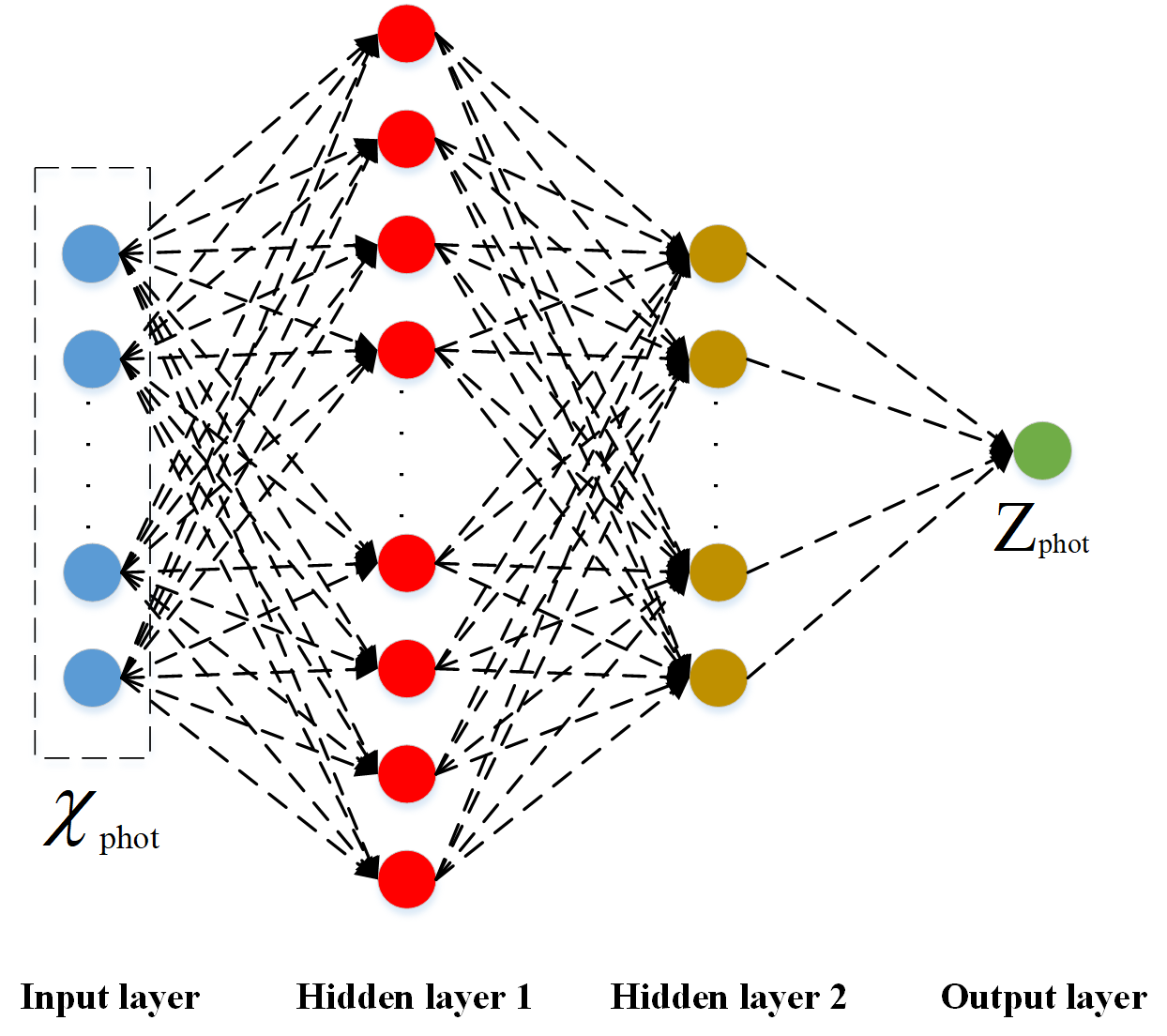}
\caption{Model structure of ANN.}
\label{f3}
\end{figure}

With only 15 input features for single-modal photometric data, there is no need to use a too deep network or complex model. We constructed a four-layer ANN model to convert the input into the feature space, and then predict the photometric redshift of quasars. Figure~\ref{f3} illustrates the model structure of the ANN constructed here.

The model is composed of one input layer, two hidden layers, and one output layer. Each circle in Figure~\ref{f3} represents one neuron. The input layer is equipped with 15 neurons corresponding to the 15 features of the input. The first hidden layer is equipped with 32 neurons, the second hidden layer with 16 neurons, and the last layer outputs the predicted redshift value. Starting from the second layer, the input of neurons in each layer is a linear combination of neurons in the previous one. Each hidden layer is added with $relu$ activation function (shown in equation~(\ref{m4})), providing nonlinear variation. 
\begin{equation}
    relu = max(0, x),
	\label{m4}
\end{equation}
where $max$ is the function of calculating the maximum value, and $x$ represents the output of the previous layer network.

It can be seen that the structure of the above model is straightforward. We have also tried to increase the depth of the model or use other complex models, such as CNN. Nonetheless, the complex model would significantly increase the training time, without necessarily better prediction results than those of the above model. Having only 15 input features, an overly complex model would thus lead to overfitting. For this, we finally chose this simple model to get better results at the lowest cost.

In the process of model training, we define the loss function as $L1$ (denoted as $loss (L1)$). It represents the average absolute error amplitude of the predicted value, as shown in equation~(\ref{m5}). Compared to the mean square error, $loss (L1)$, chosen here, has better robustness to outliers. 
\begin{equation}
    loss(L1) = \frac{\sum_{n=1}^{N}\left | z_{spec}^{n}-z_{phot}^{n} \right |}{n},
	\label{m5}
\end{equation}
where $N$ represents the total number of samples (i.e. 415,930), $n$ ranges from 1 to N, $z_{spec}^{n}$ represents the spectroscopic redshift of the $n$-th data, and $z_{phot}^{n}$ represents the predicted photometric redshift of the $n$-th data.

\subsection{Design and implementation of the PhotoRedshift-MML method}
In Section 3.1, the single-modal photometric data is used to predict photometric redshifts of quasars. As mentioned in the introduction, because of the insufficient features available for model learning, this single-modal machine learning model has encountered a bottleneck with difficulty to improve its prediction accuracy. Therefore, in Section 3.2.1, multimodal representation learning is introduced to represent data pairs (photometric data features, spectroscopic data features) in a unified feature space. The represented spectral features are used as labels to generate simulated spectral data. In 3.2.2, aiming at the problem of low redshift prediction accuracy due to few photometric features of quasars, the simulated spectra are used to assist the photometric redshift prediction of quasars. Thus, a photometric redshift prediction model based on multimodal transfer learning is constructed.

\subsubsection{Feature transformation model based on multimodal representation learning}
In some cases, only the photometric data of quasars exist due to the limitation of observation equipment. In this case, the prediction of quasar redshift can only rely on a small number of features contained in photometric data, which renders it challenging to obtain high prediction accuracy. This section aims at considering the knowledge of the real spectra as labels and transfer photometric data into spectral features so that the prediction accuracy can still be improved with the help of simulated spectra using photometric data alone. To acquire knowledge from the pre-training and achieve the conversion from photometric data to spectral features, the following two steps are required. First, in a unified feature space, the data pairs (photometric data features, spectroscopic data features) are represented, and the feature vectors of the extracted spectra are used as labels. Second, a PhotoSpecTransformer model is built to drive the photometric data to iteratively learn from the extracted spectra and convert it into simulated spectral features.

\textbf{(1) Step 1: Spectral feature representation and extraction}

We first provide a formal representation to better describe the PhotoRedshift-MML method.

We denote the spectral features extraction network as $\mathcal{F}_{0}$, its input as $\chi_{spec}$, and its output as $\chi_{spec}^{true}$, referring to equation~(\ref{m6}).
\begin{equation}
    \chi_{spec}^{true} = \mathcal{F}_{0}(\chi_{spec}, \theta _{0}),
	\label{m6}
\end{equation}
where $\theta _{0}$ is a vector composed of all the parameters in the network $\mathcal{F}_{0}$.

Then, data pairs (photometric data features, spectroscopic data features) are constructed, denoted as ($\chi_{phot}$, $\chi_{spec}^{true}$). Taking $\chi_{spec}^{true}$ as labels, we build a PhotoSpecTransformer model, named $\mathcal{F}_{1}$, referring to equation~(\ref{m7}). $\chi_{phot}$ are used to generate simulated spectral features, denoted as $\chi_{spec}^{generated}$.  
\begin{equation}
    \chi_{spec}^{generated} = \mathcal{F}_{1}(\chi_{phot}, \chi_{spec}^{true}, \theta _{1}),
	\label{m7}
\end{equation}
where $\theta _{1}$ is a vector composed of all the parameters in the model $\mathcal{F}_{1}$.

So far, we have got simulated spectral features, $\chi_{spec}^{generated}$, so that we can enrich our features to predict redshifts, and then we build the PhotoRedshift-TL model, named $\mathcal{F}_{2}$. Taking ($\chi_{phot}$, $\chi_{spec}^{generated}$) as input, we obtain the prediction results of photometric redshift, denoted as $z_{phot}$, as shown in equation~(\ref{m8}). The detail of $\mathcal{F}_{2}$ will be presented in Section 3.2.2. 
\begin{equation}
    Z_{phot} = \mathcal{F}_{2}(\chi_{phot}, \chi_{spec}^{generated}, \theta _{2}),
	\label{m8}
\end{equation}
where $\theta _{2}$ is a vector composed of all the parameters in the model $\mathcal{F}_{2}$.

\begin{figure}
\centering
\includegraphics[width=\hsize]{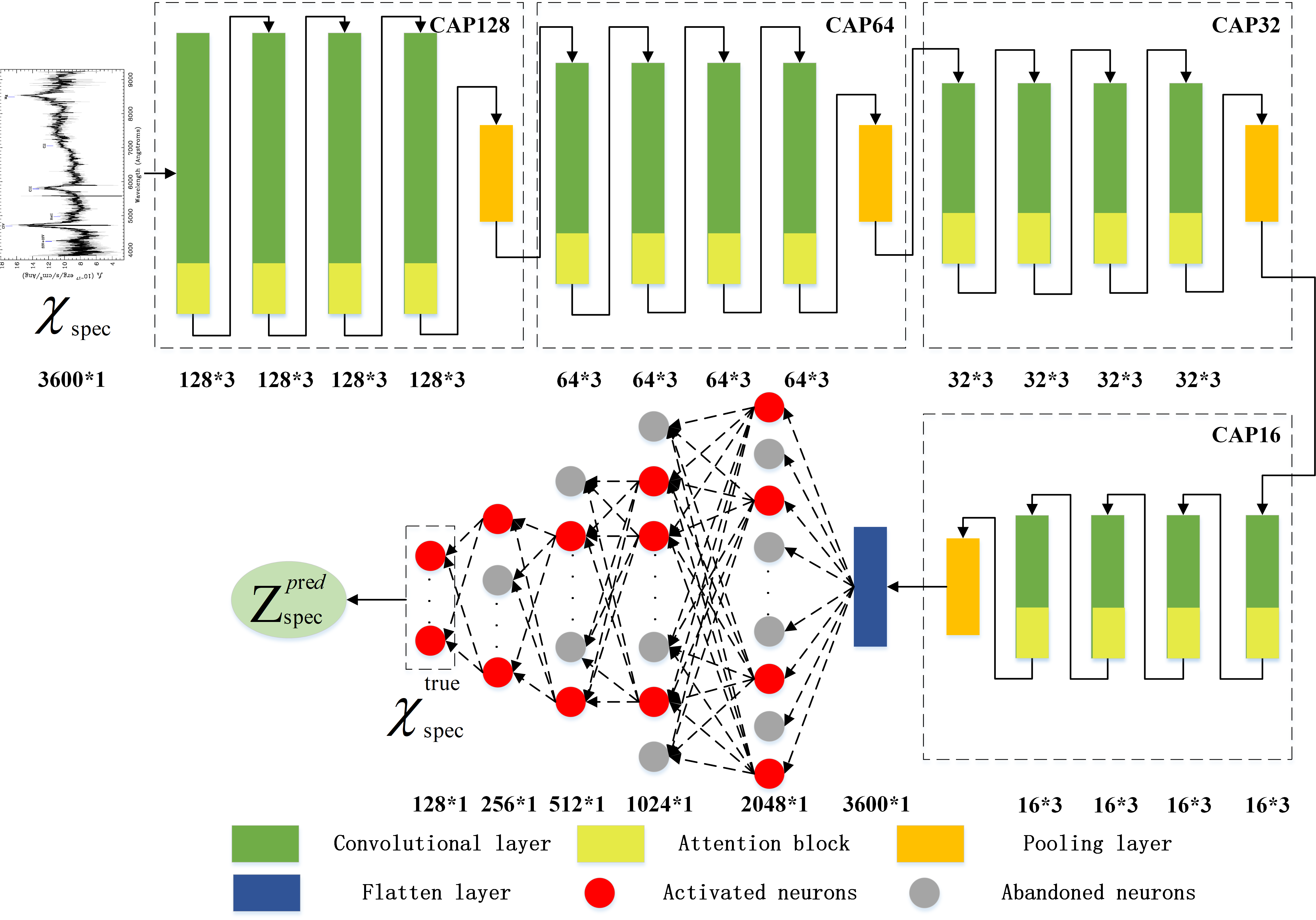}
\caption{The structure of model $\mathcal{F}_{0}$.}
\label{f4}
\end{figure}

Briefly, our PhotoRedshift-MML method is $\mathcal{F}_{2}(\mathcal{F}_{1}(\mathcal{F}_{0}))$, as generalized in Figure~\ref{f2}. The model $\mathcal{F}_{0}$ is implemented based on a convolutional neural network, as shown in Figure~\ref{f4}. It has the pre-processed 3600-dimensional spectroscopic data as input, predicting photometric redshift as the final goal, and the $loss (L1)$ function in equation~(\ref{m5}) as the objective function. After training, we retain the 128-dimensional spectral vector obtained by the model $\mathcal{F}_{0}$ as the label in step (2).

The implementation of model $\mathcal{F}_{0}$ is described in the following five parts: 

(i) One input layer; 

(ii) regard four one-dimensional convolutional layers with attention blocks and one pooling layer as a module (denoted as Convolutional-Attention-Pooling, CAP module for short). There are four modules with the same structure but different filter numbers (the number of filters in convolutional layers of each module is 128, 64, 32 and 16, respectively. Therefore, they are denoted as CAP128, CAP64, CAP32 and CAP16); 

(iii) one flatten layer; 

(iv) nine fully connected layers (the number of neurons is reduced by half from 2048 to 8 gradually, and dropout function is introduced to randomly discard half of the neurons with a probability of 0.5 to prevent overfitting);

(v) one output layer.                                                                

\textbf{(2) Step 2: PhotoSpecTransformer model construction}

\begin{figure}
\centering
\includegraphics[width=\hsize]{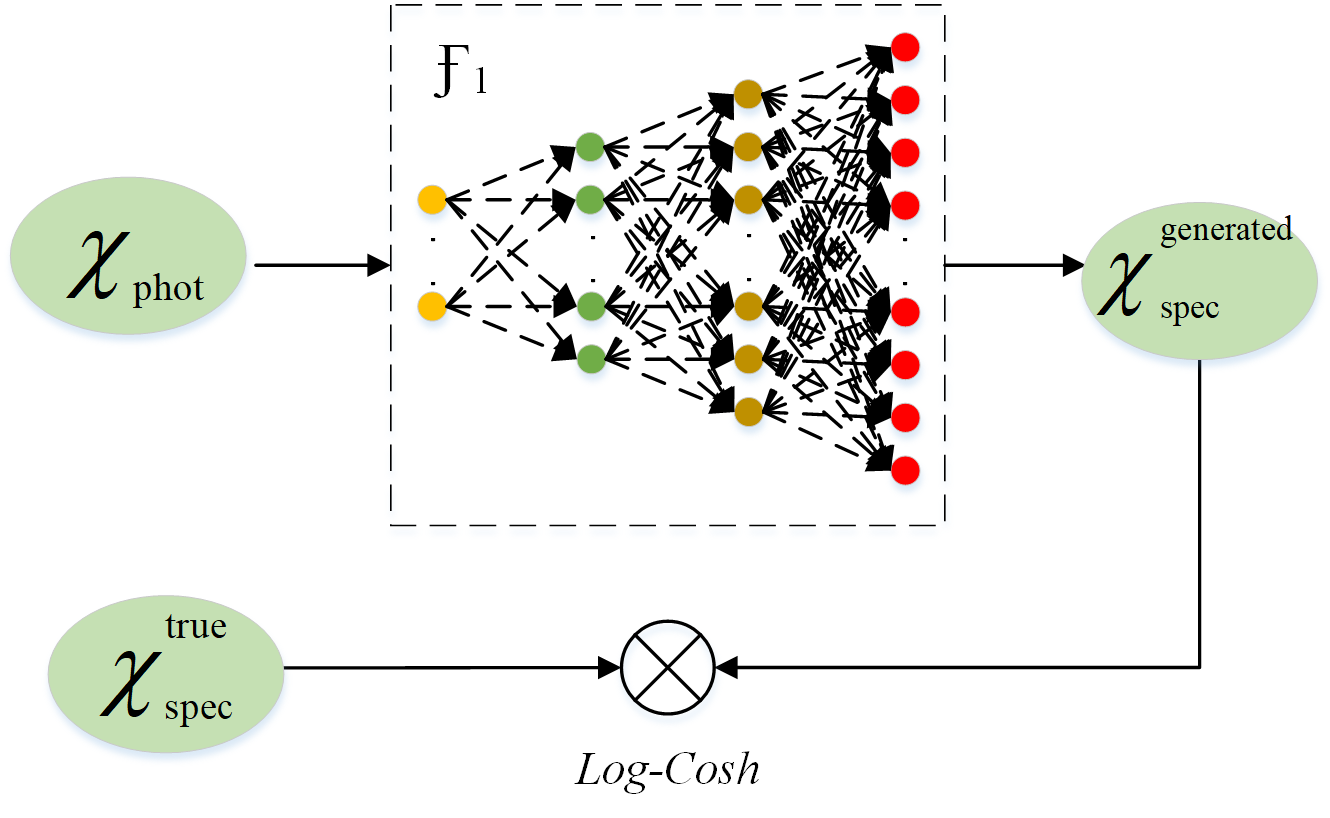}
\caption{The structure of model $\mathcal{F}_{1}$.}
\label{f5}
\end{figure}

The PhotoSpecTransformer model, i.e. $\mathcal{F}_{1}$ model, is based on ANN. As shown in Figure~\ref{f5},  the input of the model are the features of 15-dimensional photometric data, $\chi_{phot}$, and the 128-dimensional spectral feature $\chi_{spec}^{true}$ obtained in step (1), whereas the output of the model is $\chi_{spec}^{generated}$. That is, we can generate the 128-dimensional spectral feature from the 15-dimensional photometric data using the $\mathcal{F}_{1}$ model.

The implementation of model $\mathcal{F}_{1}$ is described as follows: 

(i) one input layer (including 15 neurons corresponding to the 15-dimensional input photometric data);

(ii) four hidden layers (containing 8, 16, 32 and 64 neurons, respectively). Each hidden layer is also added with $relu$ activation function (see equation~(\ref{m4})), providing nonlinear variation; 

(iii) one output layer. 

In the process of model $\mathcal{F}_{1}$ training, we iteratively optimize the cosine loss function, shown in equation~(\ref{m9}).
\begin{equation}
    Log-Cosh = \sum_{n=1}^{N}\log (\cosh (\chi_{spec}^{generated,n}-\chi_{spec}^{true,n})),
	\label{m9}
\end{equation}
where $\chi_{spec}^{true,n}$ represents the 128-dimensional true feature vector of the $n$-th spectroscopic data, $\chi_{spec}^{generated,n}$ represents the 128-dimensional simulated spectral vector.

So far, we have successfully built the PhotoSpecTransformer model, which we can further train and use. In the training stage of the model, the photometry-spectra pairs are used to iteratively optimize the parameters of the model by continuously narrowing the gap between the generated and real spectra, so that the model can generate enough similar spectral features. In the application stage of the model, the simulated spectral features can be generated adopting photometric data alone, which can assist the downstream task, i.e. quasar photometric redshift prediction.

\subsubsection{Prediction of quasar photometric redshifts based on multimodal transfer learning}

In Section 3.2.1, the 128-dimensional simulated spectral vector $\chi_{spec}^{generated,n}$, generated from the photometric data, has been obtained. Next, these simulated spectral vectors can be used to assist the 15-dimensional photometric data, $\chi_{phot}$, to more accurately predict the photometric redshift of quasars.

\begin{figure}
\centering
\includegraphics[width=\hsize]{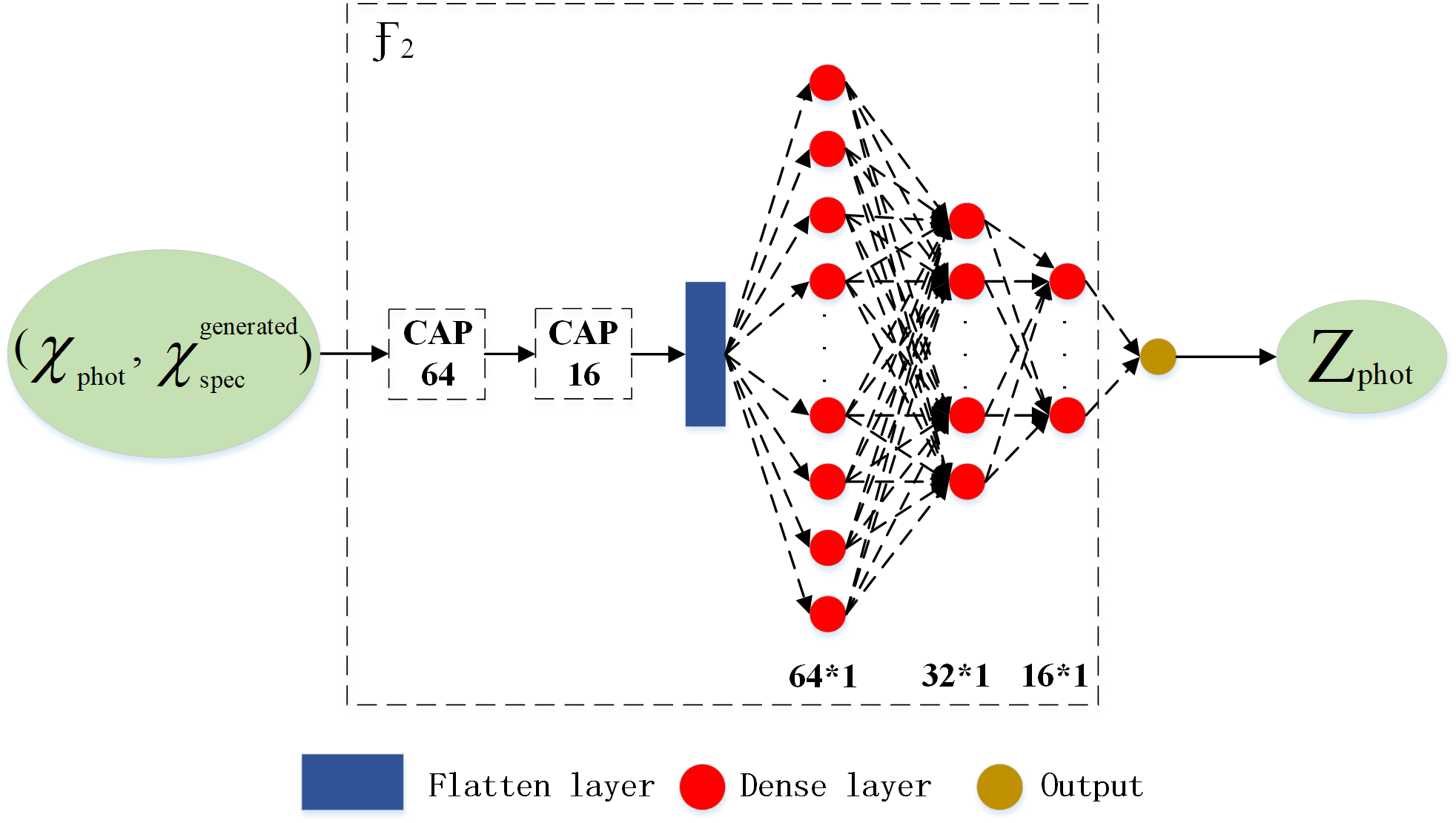}
\caption{The structure of model $\mathcal{F}_{2}$.}
\label{f6}
\end{figure}

The PhotoRedshift-TL model, i.e. model $\mathcal{F}_{2}$, similar to model $\mathcal{F}_{0}$, is built on CNN and CAP module, as shown in Figure~\ref{f6}. The specific structure will not be discussed in detail here, we only discuss the differences, which lie in the input layer and the fully connected layers. 

(i) The input of model $\mathcal{F}_{2}$ has two parts: one is Main\_input, which consists of the features of 15-dimensional photometric data $\chi_{phot}$, the other is Auxiliary\_input, which is the 128-dimensional simulated spectral features $\chi_{spec}^{generated}$ obtained in 3.2.1. The two parts of data are concatenated to form a new feature vector $V$ of 143 * 1, which is fed into CAP64 and CAP16. 

(ii) The model $\mathcal{F}_{2}$ has three fully connected layers (containing 64, 32 and 16 neurons, respectively). Since not many neurons are available in model $\mathcal{F}_{2}$, the dropout function is not used to discard neurons.

\section{Results and discussion}
In order to verify the performance of PhotoRedshift-MML, this section uses the pre-processed data in Section 2, i.e. ($\chi_{phot}$, $\chi_{spec}$). In the experiment, 415,930 pairs of data are divided into a training set and a testing set in the proportion of 8:2. The experimental results are evaluated by three evaluation metrics defined in Section 4.1, and the results are compared and analysed with baselines in Section 4.2.
\subsection{Evaluation metrics}
We use six metrics to verify the prediction performance of the model. For the prediction of galaxy redshift, the commonly used statistical mean square error, i.e. MSE, is selected as the evaluation metric. The calculation method is shown in equation~(\ref{m10}). For the prediction of quasar redshift \citep[][]{zhang2013estimating}, we introduce classical |$\Delta$z| and Root Mean Square (RMS) of |$\Delta$z|, defined in equation~(\ref{m11}) and equation~(\ref{m12}), respectively. To further measure the quality from different perspectives, we also calculate the fraction of outliers \citep[][]{cavuoti2012photometric}, defined in equation~(\ref{m13}), normalized median absolute deviation (NMAD) \citep[][]{ilbert2008cosmos}, defined in equation~(\ref{m14}) and bias \citep[][]{cavuoti2012photometric}, defined as mean($\Delta$z) after excluding outliers.

\begin{equation}
    MSE = \frac{\sum_{n=1}^{N}(z_{spec}^{n}-z_{phot}^{n})^{2}}{N},
	\label{m10}
\end{equation}

\begin{equation}
    \left | \Delta z \right | = \left | \frac{z_{phot}-z_{spec}}{1+z_{spec}} \right |,
	\label{m11}
\end{equation}

\begin{equation}
    RMS = \sqrt{\frac{\sum_{n=1}^{N}\left | \Delta z \right |_{n}^{2}}{N}},
	\label{m12}
\end{equation}

\begin{equation}
    \left | \frac{z_{phot}-z_{spec}}{1+z_{spec}} \right | > 0.15,
	\label{m13}
\end{equation}

\begin{equation}
    \sigma_{NMAD} = 1.48 * median(\left | \Delta z \right |),
	\label{m14}
\end{equation}
where $z_{spec}^{n}$ represents the spectroscopic redshift of the $n$-th data, $z_{phot}^{n}$ represents the predicted photometric redshift of the $n$-th data, |$\Delta$z|$_{n}$ represents the |$\Delta$z| of the $n$-th data.

\begin{figure}
\centering
\subfigure[]{  
\includegraphics[width=\hsize]{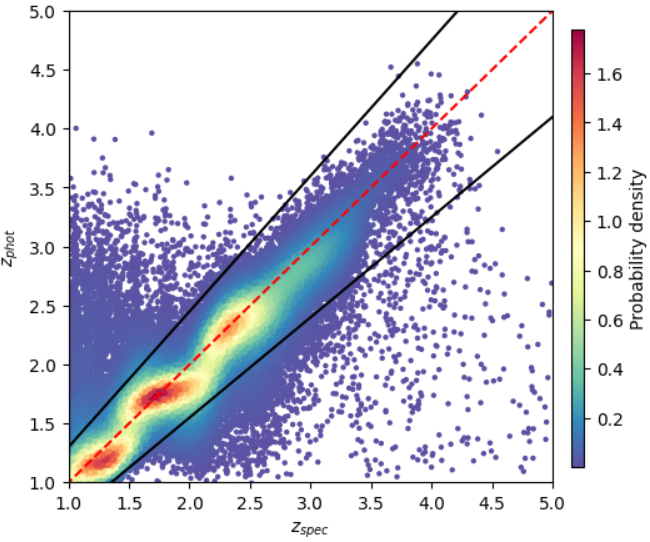}  
}
\quad
\subfigure[]{  
\includegraphics[width=\hsize]{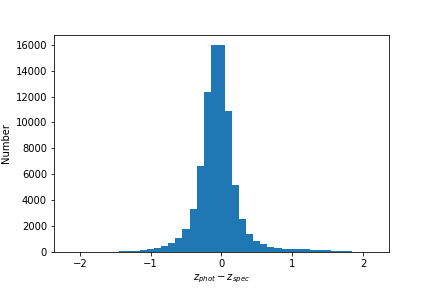}
}
\caption{Prediction results of photometric redshift based on ANN.}    
\label{f7}       
\end{figure}

\begin{figure}
\centering
\subfigure[]{  
\includegraphics[width=\hsize]{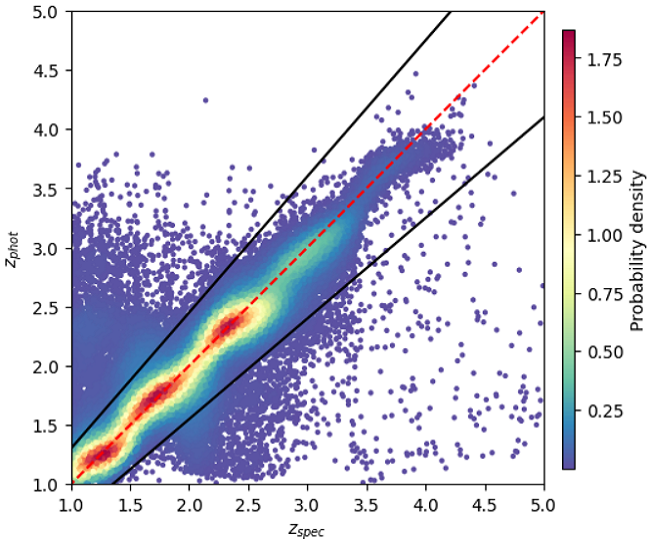}  
}
\quad
\subfigure[]{  
\includegraphics[width=\hsize]{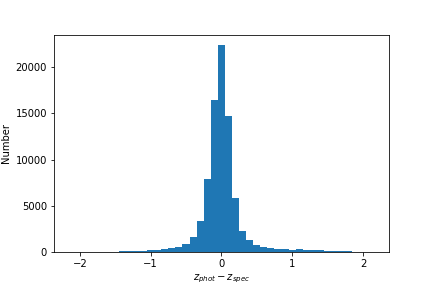}
}
\caption{Prediction results of photometric redshift based on PhotoRedshift-MML.}    
\label{f8}       
\end{figure}

\subsection{Comparison and analysis}
The experimental environment of this paper is as follows. We use an Intel Xeon E5-2690*2.6GHz CPU and an Nvidia Tesla K40 12GB GPU. The software environment includes Python 3.5, Keras 2.3.1, Numpy 1.16.2, Matplotlib 3.0.3 and Scikit\_learn 0.19.1.

In terms of parameters selection, the batch\_size is set to 512, the optimizer selects Adam, and the learning rate is 0.001.

\textbf{(1) Experiment 1: verifying the generality}

The generality of the method is reflected in that our model can also be used to perform other downstream tasks, such as predicting the photometric redshifts of galaxies with high accuracy, either using the model presented in Section 3.1 or Section 3.2. Considering that the MSE of galaxy photometric redshift prediction obtained by the model described in Section 3.1 is already low, which reaches 0.0011, there is no need to use multimodal machine learning to increase accuracy. Taking \citet{mu2020photometric} as the Baseline 1, using the same galaxy data set, we use ANN proposed in Section 3.1 instead of CNN to predict the photometric redshift of galaxy data and calculate the MSE of the prediction results. It is found that the prediction accuracy is equivalent (even slightly better) to that in Baseline 1, as shown in Table~\ref{tab1}. This is because when there are few input characteristics of the model, their CNN is not the optimal choice. See Table~\ref{tab1} for the specific comparison results. As for the generality of the PhotoRedshift-TL model, it can be effectively used as long as the quasar data pairs are replaced with the data pairs of other tasks. 

\begin{table}
\centering
\caption{Comparison of Galaxy photometric redshift prediction results between Baseline 1 and our method.}
	\label{tab1}
    \begin{tabular}{lccr}
        \\\hline
        \multicolumn{1}{c}{Data Set}    & \multicolumn{1}{c}{MSE}
        \\\hline
        \multicolumn{2}{c}{Results of Baseline 1}          
        \\\hline
        SDSS DR13(early-type galaxy)         & 0.0014                  \\
        SDSS DR13(late-type galaxy)          & 0.0019                
        \\\hline
        \multicolumn{2}{c}{\textbf{Our results}} &                       
        \\\hline
        SDSS DR13(galaxy)                    & 0.0011        
        \\\hline
    \end{tabular}
\end{table}

\textbf{(2) Experiment 2: verifying the effectiveness and advantage}

For the photometric redshift prediction of quasars, we draw the density diagram and prediction error distribution diagram of photometric redshift $z_{phot}$ as predicted results versus spectroscopic redshift $z_{spec}$ as true labels of all data on the testing set, as shown in Figure~\ref{f7} and Figure~\ref{f8}.



The abscissa of Figure~\ref{f7} (a) and Figure~\ref{f8} (a) is $z_{spec}$ with the ordinate being $z_{phot}$. The prediction results are drawn on the left part of the figures, and the corresponding colour bar is described on the right part of figures. As for the colour bar, the more points, the redder the colour, whereas for less points, the bluer is the colour. Figure~\ref{f7} (b) and figure~\ref{f8} (b) show the statistical number of samples with $z_{phot}$-$z_{spec}$ in different intervals. The abscissa is $z_{phot}$-$z_{spec}$, the bin is 0.1, the middle bin ranges from -0.05 to +0.05, and the ordinate is the number of samples. Comparing Figure~\ref{f7} (a) to Figure~\ref{f8} (a) shows that the error of redshift prediction based on PhotoRedshift-MML is significantly reduced and the data is more concentrated near the diagonal, indicating that the predicted value is closer to the real one. In other words, it is effective to add the simulated spectra obtained through transfer learning as an auxiliary input. For example, the proportion of data with |$\Delta$z| within 0.1 increases from 80.41\% to 84.45\%. Comparing Figure~\ref{f7} (b) to Figure~\ref{f8} (b) shows that the amount of data with a smaller prediction error of $\pm5\%$ has increased by a quarter from 17502 to 21780. In comparison, the amount of data with a larger prediction error has been significantly reduced. There are two reasons for the above results. One is that our PhotoRedshift-MML method can make good use of photometric data and simulated spectral vectors. The other is that the number of features increases from 15 to 143 by adding the simulated spectral vector close to the real spectral vector.

We calculate the percentages of different |$\Delta$z| intervals and RMS of |$\Delta$z| of the above experimental results, by taking \citet{zhang2013estimating} as the Baseline 2. The results are shown in Table~\ref{tab2}.

\begin{table*}
\centerline{}
    \caption{\centering Comparison of quasar photometric redshift prediction results between Baseline 2 and our method.}
	\label{tab2}
    \begin{tabular}{|c|c|c|c|c|c|}
        \\\hline
        Data Set & Input Pattern & |$\Delta$z|\textless{}0.1(\%) & |$\Delta$z|\textless{}0.2(\%) & |$\Delta$z|\textless{}0.3(\%) & RMS of |$\Delta$z| 
        \\\hline
        \multicolumn{6}{c}{Results of Baseline 2} 
        \\\hline
        SDSS DR7 & 4C, r & 78.63±0.23 & 85.70±0.27 & 87.09±0.23 & 0.259±0.003 \\
        
        SDSS DR7-WISE & 6C, i & 88.64±0.25 & 96.38±0.15 & 97.63±0.10 & 0.117±0.004 \\
        \hline
        \multicolumn{6}{c}{\textbf{Our results}} 
        \\\hline
        SDSS DR17 & $\chi_{phot}$ & 80.41 & 92.60 & 96.06 & 0.1332 \\
        SDSS DR17 & ($\chi_{phot}$, $\chi_{spec}^{generated}$) & 84.45 & 93.59 & 96.43 & 0.1235\\
        SDSS DR8 & $\chi_{phot}$ & 77.13 & 94.47 & 97.13 & 0.1174 \\
        SDSS DR8 & ($\chi_{phot}$, $\chi_{spec}^{generated}$) & 90.78 & 95.62 & 97.35 & 0.1046
        \\\hline
    \end{tabular}
\end{table*}

\citet{zhang2013estimating} uses different datasets from SDSS DR7 and SDSS DR7-WISE. Their experimental results show that the more data sets used, the richer the information contained and the higher the accuracy of predicting photometric redshift. When using SDSS data alone, the proportion of data with |$\Delta$z| < 0.1 accounts for 78.63\%. After adding WISE, the proportion of data with |$\Delta$z| < 0.1 increases to 88.64\%. Correspondingly, RMS of |$\Delta$z| decreases from 0.259 to 0.117.

As for our results of photometric redshift estimation, we conducted experiments on two data sets, SDSS DR8 and SDSS DR17. When using the photometric data of SDSS DR8 and SDSS DR17 alone, the proportion of data with |$\Delta$z| < 0.1 accounts for 77.13\% and 80.41\%, respectively, which are similar to 78.63\% in Baseline 2. However, the proportion of data with |$\Delta$z| < 0.2 accounts for 94.47\% and 92.60\%, respectively, which are much higher than that of Baseline2. Moreover, the proportion of data with |$\Delta$z| < 0.3 accounts for 97.13\% and 96.06\%, respectively, which are much higher than that of Baseline 2. Our RMS of |$\Delta$z| is 0.1174 on SDSS DR8 and 0.1332 on SDSS DR17, whereas that in Baseline2 is 0.259. this indicates that the overall performance of our ANN model is much better than Baseline2. After adding the generated simulated spectral vector as an auxiliary input on SDSS DR8, the proportion of data with |$\Delta$z| within 0.1, obtained by our PhotoRedshift-MML method, increases to 90.78\%, which is about 13\% higher than that of a single modal method. Whereas on SDSS DR17, the proportion increases by 4.04\% (i.e. proportion of data reaches 84.45\%). The proportion of data with |$\Delta$z| < 0.2 and |$\Delta$z| < 0.3 also increase, although they are not very significant because they are already very high in single modal data. RMS of |$\Delta$z| decreases from 0.1174 to 0.1046 on SDSS DR8 and from 0.1332 to 0.1235 on SDSS DR17, which shows that the photometric redshift predicted by multimodal transfer learning is more accurate than that predicted by using photometry alone on the whole. Throughout the four statistical results of PhotoRedshift-MML, the effect of using photometric data fused with 128-dimensional simulated spectral features from $\mathcal{F}_{1}$ model can be comparable to that of using SDSS-WISE.

\
\textbf{(3) Quality assessment and comparisons.}

In order to evaluate the prediction results from different perspectives, we also calculate three commonly used statistical metrics: the fraction of outliers, NMAD, and bias. The results can be seen in Table~\ref{tab3}. 

\begin{table}
\centering
\caption{Comparison between single modal and multi-modal method on SDSS DR17.}
	\label{tab3}
    \begin{tabular}{|c|c|c|c|}
        \\\hline
        Input Pattern    &
        Outlier Fraction       &
        $\sigma_{NMAD}$    &
        Bias 
        \\\hline
        $\chi_{phot}$         & 0.1050  &  0.0732  & -0.0192    \\
        ($\chi_{phot}$, $\chi_{spec}^{generated}$) &   0.0905 &  0.0520  &  -0.0042               
        \\\hline
    \end{tabular}

\end{table}

Typically, outliers are those objects who meet equation~(\ref{m13}), which is a quality control metric \citep[][]{cunha2022photometric}. On SDSS DR17, with single modal method, outliers account for 10.50\%, whereas with PhotoRedshift-MML method, outliers account for 9.05\%. The outlier fraction reduces by 1.45\% with the auxiliary of generated spectral features.

The NMAD, defined in equation~(\ref{m14}), provides a measure of the variability in the sample. This dispersion estimate is less sensitive to outliers \citep[][]{brammer2008eazy}. The results of $\sigma_{NMAD}$ are 0.0732 and 0.0520 for single modal method and multi-modal method, respectively.

The bias shows the average separation between prediction and true values \citep[][]{li2022photometric}. It is quite small in both methods. Particularly, the bias in PhotoRedshift-MML method is only -0.0042, which is 0.015 closer to zero than the single modal method. The closer the bias is to zero, the smaller the systematic bias is, between photometric and spectroscopic redshifts \citep[][]{dahlen2013critical}.

\section{Conclusion}
Although machine learning is widely used in estimating quasar photometric redshifts, we believe that the present work provides a first development of a quasar photometric redshift prediction method based on multimodal machine learning in this field. Furthermore, we demonstrate the effectiveness and advantage of our PhotoRedshift-MML by conducting extensive experiments. Specifically, in the pre-training stage, the photometric data are used to learn the spectral feature representation, and then the photometry-spectra feature transformation model is constructed. The trained model can make the photometric data generate very similar spectral feature vectors. In the downstream task, only photometric data is available, the prediction accuracy of photometric redshift can be significantly improved with the assistance of simulated spectral feature vectors.

In order to verify the effectiveness, advantage and generality of our method, we used the data from SDSS DR8 and DR17 to form an even pair (photometric data, spectroscopic data) for the same quasar observation source. The major advantage of PhotoRedshift-MML is that we first learn knowledge of the real spectra, and then convert photometric data to spectral features by transfer learning. This greatly reduces the reliance of collecting spectroscopic data to get more input features.

In terms of effectiveness of the method, we used SDSS DR8 and DR17 photometric data alone for single-modal redshift prediction. Due to the fact that the input features are less and small, we build a 4-layer ANN network. The predicted quasar photometric redshift of |$\Delta$z| < 0.1 accounts for 77.13\% and 80.41\% of the total test samples from SDSS DR8 and DR17 respectively, which are close to 78.63\% obtained by \citet{zhang2013estimating}. However, the proportion of data with |$\Delta$z| < 0.2 and |$\Delta$z| < 0.3, both vastly increased compared to that in \citet{zhang2013estimating}. The RMS of |$\Delta$z| has also decreased greatly. Therefore, our method has certain effectiveness.

In terms of advantage of the method, we used multimodal data to predict the redshift. After adding the additional features of the simulated spectrum, the proportion of data with |$\Delta$z| < 0.1 accounts for 90.78\% on SDSS DR8 and 84.45\% on SDSS DR17, increasing by 13.65\% and 4.04\% respectively compared to the single modal case. In addition, the RMS of |$\Delta$z| is 0.0128 and 0.0097 lower than that in the single modal. These results are comparable to the prediction results obtained by using the two data sets of SDSS-WISE in \citet{zhang2013estimating}, which proves that our method is superior.

In terms of its generality, our model would only require minor changes to perform other tasks, such as galaxy classification and galaxy redshift prediction. For example, the MSE of the prediction result with the method in Section 3.1 can reach 0.0011, which is equivalent to, or even more accurate, than the results of 0.0014 and 0.0019 in \citet{mu2020photometric}. Therefore, our method also has certain generality.

In conclusion, our PhotoRedshift-MML can achieve the best RMS of |$\Delta$z| and significantly outperform the previously published state-of-the-art work in a single modal since it can enrich the characteristic features by exploiting another simulated modal data. In future research works, we plan to fuse the data of other sky survey plans, such as UKIDSS and WISE, which is expected to further improve the accuracy of multimodal prediction of redshift with the help of more observation data. Furthermore, we believe that the method of multimodal machine learning is a general method, and plan to apply it to other tasks, e.g. photometric redshifts estimation of galaxies and classification of celestial objects.

\section*{Acknowledgements}

This work is supported by the science research grants from the National Science Foundation of China (No. U1931209) and the China Manned Space Project with NO.CMS-CSST-2021-B05. The authors are also highly grateful for the constructive suggestions given by Jiali Deng.

\section*{Data Availability}

The data underlying the research results is available in SDSS DR17, on http://skyserver.sdss.org/CasJobs/ and https://dr17.sdss.org/optical/spectrum/search for photometric data and spectroscopic data, respectively.



\bibliographystyle{mnras}
\bibliography{ref} 



\appendix

\section{SQL query}

select $a.plate$, $a.mjd$, $a.fiberID$, $a.specObjID$, $a.bestObjID$, $a.class$, $a.z$, $a.zErr$, $b.objID$, $b.psfMag\_u$, $b.psfMag\_g$, $b.psfMag\_r$, $b.psfMag\_i$, $b.psfMag\_z$, $b.extinction\_u$, $b.extinction\_g$, $b.extinction\_r$, $b.extinction\_i$, $b.extinction\_z$, $b.psfMagErr\_u$, $b.psfMagErr\_g$, $b.psfMagErr\_r$, $b.psfMagErr\_i$, $b.psfMagErr\_z$

into $mydb.dr17psf\_ugriz\_SNR$

from $dr17.SpecObjAll$ as $a$

join $dr17.PhotoObjAll$ as $b$

on $a.bestObjID = b.objID$

where $a.class$ = $'QSO'$ and $a.z$ >= 1 and $a.z$ <= 5 and $a.zErr$ < 0.001 and $a.snMedian\_u$ < 10 and $a.snMedian\_g$ < 10 and $a.snMedian\_r$ < 10 and $a.snMedian\_i$ < 10 and $a.snMedian\_z$ < 10 and $b.psfMag\_g$ >= 18 and $b.psfMag\_g$ <= 22 and $b.petroRad\_r$<5


\bsp	
\label{lastpage}
\end{document}